\documentclass[preprint,showpacs,preprintnumbers,amsmath,amssymb]{revtex4}
\newcommand{\ra}{\rangle}
\newcommand{\la}{\langle}
\usepackage{graphicx}

\begin{document}

\title{Anomalous finite-size effects 
and canonical asymptotic behaviors for the mean-squared gyration radius of 
 Gaussian random knots}
 
\author{Miyuki K. Shimamura}
 \email{smiyuki@exp.t.u-tokyo.ac.jp}
\address{Department of Advanced Materials Science, 
Graduate School of Frontier Sciences, University of Tokyo,
7-3-1 Hongo, Bunkyo-ku, Tokyo 113-8656, Japan}
\author{Tetsuo Deguchi}
% \email{deguchi@phys.ocha.ac.jp}
\address{
Department of Physics, Faculty of Science,
 Ochanomizu University,
2-1-1 Ohtsuka, Bunkyo-ku, Tokyo 112-8610, Japan}

\begin{abstract} 
Anomalously  strong finite-size effects have been observed 
for the mean square radius  of gyration $  R^2_K $ 
of Gaussian random polygons  with a fixed knot $K$  
as a function of the number $N$ of polygonal nodes.  
 Through  computer simulations with $N < 2000$, 
 we find  for several  knots 
that the gyration radius $ R^2_K$  
 can be approximated by a power law: 
$ R^2_K \sim  \, N^{2 \nu_{K}^{\rm eff}} $,  
  where the effective exponents $\nu_{K}^{\rm eff}$ for the knots are   
   larger than 0.5 and less than 0.6. 
  A crossover occurs  for the gyration radius of the trivial knot, 
 when $N$ is roughly equal 
 to the  characteristic length $N_c$ of random knotting.    
 For the asymptotic behavior of $R^2_K$,  
 however,  we find  that  it is  consistent with  
  the standard one with the scaling exponent 0.5.      
 Thus, although  the strong finite-size effects of $R^2_K$ 
 remain effective  for extremely large values of $N$, 
 they can be matched  with the  
 asymptotic behavior of random walks.   
\end{abstract}

\pacs{36.20.-r, 61.41.+e, 05.40.Fb}
\maketitle

\newpage
\section{Introduction}

\par 
Topological effects on  statistical and dynamical properties
of ring polymers should  be quite nontrivial.  
The topological state of a ring polymer  is 
described by a knot type, 
and it is invariant after its synthesis.   
Knotted ring polymers or knotted DNAs have been discussed 
theoretically since the 1960s \cite{Delbruck,Frisch-Wasserman},  
 and recently they are  synthesized in several experiments 
 \cite{Rybenkov-Shaw,StasiakJMB}.  
Various topological effects on ring 
polymers have been explicitly 
 studied through  numerical simulations  
of random polygons under topological constraints  
\cite{Vologodskii1,desCloizeaux,Michels,DeguchiJKTR}. 
 However, many questions still remain unsolved even on the average size of 
a knotted ring polymer in solution, which should be     
the most fundamental quantity in the physics of ring polymers.  

\par 
Recently, it has been suggested  \cite{desCloizeaux-Let,Deutsch,Grosberg} that 
the average size $R_{triv}$  of a random polygon with the trivial knot 
 should  scale as $N^{\nu}$ with respect to 
the number $N$ of polygonal nodes, where the exponent $\nu$ is 
given by the exponent $\nu_{\rm SAW}$ of the asymptotic scaling-behavior 
of the the self-avoiding walk (SAW) where   
$\nu_{\rm SAW} \approx 0.588$. 
Here we remark that 
random polygons correspond to ring polymers with no excluded volume.  
There has also been  a conjecture  \cite{desCloizeaux-Let} that 
 the effect of topology on the average size of ring polymers 
 could play a similar role as the excluded-volume effect,  
since the topological constraint 
should effectively lead to  an entropic repulsion 
among the monomer segments.   
Here, we note that the trivial knot (or the unknot) 
is such a knot that is equivalent to an unknotted circle.

\par  
The conjecture on the entropic repulsion  
 seems to be  quite interesting,  
and the anomalous scaling behavior with an enhanced 
exponent should be effective at least for some numerical simulations. 
 However, it is not trivial to understand 
the consequence that the scaling exponent should be enhanced and  
given by that of SAW.  
First of all, under no topological constraint, 
 the average size $R$ of a random polygon 
scales as $N^{\nu_{\rm RW}}$ where $\nu_{\rm RW}=0.5$. 
Furthermore, it is not clear whether the scaling exponents 
of nontrivial knots should be enhanced similarly 
as that of the trivial knot. 
If they might have the same exponent of SAW, then why the 
total average over all knot types 
can have the exponent  $\nu_{\rm RW}$ of random walk ?  
The purpose of this paper is to discuss these questions 
explicitly through numerical simulations.   
We  evaluate  the mean square radius of gyration 
$R_K^2 $  of  Gaussian random polygons with a fixed knot type $K$.   
Discussing the $N$-dependence of $R_K^2$ 
for several different knot types,   
we show that the anomalous scaling behavior 
should be considered as a strong finite-size effect 
which could be valid  for very large values of $N$ such as 2000. 

\par 
We now review some relevant results on the topological effects 
of ring polymers. 
Let us take a model of  random polygons of $N$ nodes \cite{desCloizeaux,Michels}, 
which describes  ring polymers consisting of $N$ Kuhn units 
at the theta condition.  
We denote by $P_K(N)$ the probability  of a given configuration of 
the random polygon of $N$ nodes having a fixed  knot type $K$. 
For the trivial knot, it was numerically shown \cite{Michels,DeguchiJKTR,Koniaris} that 
the probability is given by an exponential function of $N$: $P_{triv}(N)= \exp(-N/N_c)$.
For nontrivial knots, the probability is well described by the following function of $N$:  
$P_K(N)=C_K \, (N/N_c)^{m(K)} \exp( -N/N_c)$, 
where we call $N_c$ and $m(K)$  the characteristic length of random knotting and 
the topological exponent of the knot, respectively \cite{DeguchiRevE}. 
The value of $N_c$ is model-dependent, 
and is roughly given by 340 for the Gaussian random polygon
\cite{DeguchiJKTR,DeguchiRevE}. We remark that the number $N_c$ 
is important in the analysis of topological effects 
with the blob picture \cite{Grosberg}. 

\par 
The mean-squared gyration radius $R_K^2$
 under the topological constraint of knot $K$ 
 has been  discussed for some models of  self-avoiding 
polygons (SAPs) in Refs. \cite{LeBret,Chen,Klenin,Janse1991,turuD,Orlandini,PRE}.
In the lattice model, it is shown that the asymptotic behavior of 
$R_K^2$ is consistent with that of the RG theory where 
 in the large $N$ limit  the  ratio  $R_K^2/R^2 $ comes close to 1.0 for any knot. 
However, for the cylinder model of SAPs \cite{PLA}, it is found \cite{PRE} that 
the limit of the ratio depends on the 
cylinder radius which controls the excluded-volume. 
For a lattice model of random polygons \cite{Brinke},  
$R_K^2$ has  been evaluated 
for the trivial and trefoil knots with small polygons of $N < 200$.

\section{Methods of simulations }

We now introduce the Gaussian random polygon \cite{desCloizeaux}. 
Let ${\vec X}_1, \cdots, {\vec X}_N$ denote the position vectors of the
nodes of a configuration of the  Gaussian random polygon of $N$ segments, 
and ${\vec u}_1, \cdots, {\vec u}_N$ 
 the jump vectors such that ${\vec u}_j={\vec X}_j-{\vec X}_{j-1}$ 
 for $j=1, \ldots, N$. 
Then, the Gaussian random polygon has the 
following  distribution function of the jump vectors: 
$P({\vec u}_1,\cdots,{\vec u}_N)=Const. 
\times \exp (-({\vec u}_1^2+ \cdots+{\vec u}_N^2)/2) 
\, \delta({\vec u}_1+\cdots +{\vec u}_N)$ . 
Here $\delta({\vec x})$ denote Dirac's delta function in three dimensions.
We construct $M$ configurations of 
the  Gaussian random polygon with $N$ nodes 
by the conditional probability distribution \cite{desCloizeaux}
\begin{equation}
P({\vec u}_j; {\vec u}_1,\cdots, {\vec u}_{j-1})=(2 \pi)^{-3/2} 
 \exp \Biggl( -\frac{N-j+1}{2(N-j)} 
 ( {\vec u}_j + {\frac{{\vec X}_{j-1} }{N-j+1}} )^2 \Biggr)
\end{equation}

\par 
Let us consider $M$ samples of randomly constructed polygons 
with $N$ nodes. We denote by $M_K$ the number of polygons with  knot $K$. 
We determine the number $M_K$ by enumerating 
 such polygons in the set of $M$ polygons 
that have the same set of values 
of the following two invariants for  knot $K$: 
the determinant of knot $\Delta_K(-1)$ 
and the Vassiliev invariant  $v_2(K)$ of the second degree 
\cite{DeguchiPLA,Polyak}.

\section{Anomalous finite-size behaviors of $R_K^2$}

\par 
 Let us discuss the numerical data of our simulations. 
 For Gaussian random polygons 
 with several different numbers $N$ up to 1900,  
 numerical estimates of $R^2$ and $R_K^2$ have been obtained 
 for the four knots: the trivial, trefoil ($3_1$) and figure-eight 
 ($4_1$) knots, and the composite knot consisting of two trefoil knots 
  ($3_1 \sharp 3_1$).  Here, we take $M=10^5$ in all the  simulations.  
  
\par 
We recall that under no topological constraint,
 the mean square radius of gyration $R^2$ of a  random polygon with $N$ nodes 
 is defined by $R^2= \sum_{n,m=1}^{N} \la (\vec{R_n}-\vec{R_m})^2 \ra/2N^2$.
Here ${\vec R}_n$ is the position vector of the $n$-th node and 
the symbol $\la \cdot \ra$ denotes the  statistical average,   
 which is given by the average over $M$  polygons in the simulations. 
For a  knot $K$, the quantity $R_K^2$ is given by 
 $R_K^2=\sum_{i=1}^{M_K}R_{K,i}^2/M_K$, 
where $R_{K,i}^2$ denotes the gyration radius of the $i$-th Gaussian random polygon
that has the knot type $K$, in the set of $M$ polygons. 
In terms of $R_K^2$, $R^2$ is given by  $R^2=\sum_{K}M_K R_K^2/M$.

\par 
In Fig. 1, the graphs of  the ratio $R_K^2/R^2$ against the the number $N$ 
are depicted  for the trivial, trefoil and figure-eight knots. 
 The fitting curves in Fig. 1 are given by 
 $R_K^2/R^2= \Gamma_K N^{2 \Delta \nu_K^{\rm eff}}$.  
The best estimates of the fitting parameters are given in Table 1.

\par 
We see in Fig. 1  that the ratio $R_K^2/R^2$ is not constant with respect to $N$.  
The ratio $R_K^2/R^2$ increases monotonically for all the knots. 
 For the trivial knot, the ratio $R_{triv}^2/R^2$ is always larger than 1.0.
On the other hand, for the other three knots 
($K$=$3_1$, $4_1$, $3_1 \sharp 3_1$), the ratio 
 $R_K^2/R^2$ is smaller than 1.0 when $N$ is small. 
However, the ratio $R_K^2/R^2$ becomes larger than 1.0 when 
  $N$ is large.
Thus, the  topological constraint gives an effective swelling 
for the  case of large $N$. 
Here we have a conjecture that for any nontrivial knot $K$,   
 $R_K^2$ should be larger than $R^2$  if $N$ is large enough. 

\par 
Let us consider  the plot of the trivial knot shown in Fig. 1.
There is a nontrivial finite-size behavior: 
the ratio $R_{triv}^2/R^2$ is constant wit respect to $N$ 
 when $N$ is very small,  
while  when  $N > N_c$, it can be approximated by a scaling behavior 
as $R_{triv}^2/R^2 \sim N^{2\nu_{triv}^{\rm eff}}$, at least up to $N=2000$. 
This crossover phenomenon should  be 
 consistent with the recent theory given  by Grosberg \cite{Grosberg}. 
However, the effective exponent $\nu_{triv}^{\rm eff}$ 
  is much smaller than 
the exponent $\nu_{\rm SAW}$ with respect to the errors. 
In fact, we have the numerical estimate: 
 $\nu_{triv}^{\rm eff} \approx 0.545$.

\par 
 For the case of nontrivial knots ($3_1$, $4_1$, $3_1 \sharp 3_1$), 
the ratio $R_K^2/R^2$ is well approximated by the power law: 
$R_K^2/R^2 \approx \Gamma_K N^{2 \Delta \nu_K^{\rm eff}}$ 
for the range from $N=100$ to $N=2000$. 
Furthermore, there is no crossover for the nontrivial knots:  
we do not find any change in the  slope of the graphs 
near $N \sim N_c$.   It is also remarkable 
 from Table 1 that the exponent  $\Delta \nu_K^{\rm eff}$ 
 strongly  depends on the knot type. In particular, 
 the effective scaling exponent 
  of the composite knot  $3_1 \# 3_1$ 
   is almost as large as the exponent 
   $\nu_{\rm SAW}$, while that of the trefoil knot is given by   
  0.561, which is rather smaller  than  $\nu_{\rm SAW}$ with 
  respect to the errors.

\par 
Let us consider the three fitting lines of Fig. 1. Then, we see 
that the three lines become very close to each other at around $N=2000$.  
Furthermore,  it is suggested from the simulations  that 
when $N$ becomes close to 2000,  
the values of  $R_K^2$ for the four knots 
($K$=trivial, $3_1$, $4_1$ and $3_1\sharp 3_1$) 
should become  almost equal to each other. 
 In fact, up to $N=1900$,  the values of $R_K^2$  
 for the nontrivial knots  are 
always smaller than or equal to that of $R_{triv}^2$ in our simulations. 
If the power-law approximation might be valid  
 also for $N > 2000$, then $R_{triv}^2$ would become  
much smaller than $R_{K}^2$ of the three nontrivial knots for large $N$ 
and it would be inconsistent with the numerical  results obtained so far.   
 Thus, we may conclude that the approximation of $R_K^2$ with the power law 
 should be valid only when $N < 2000$. 
 Therefore, in order to study  the $N$-dependence  
 of $R_K^2$ for  $N > 2000$, 
 we  need another independent analysis.

\section{Asymptotic behavior of $R_K^2$ }

Let us discuss the  asymptotic behavior of $R_K^2$ . 
When $N$ is very large, we may assume the following expansion: 
$R_K^2=A_K N^{2\nu_K} (1+ B_K N ^{-\Delta} +O(1/N))$.
It is consistent with renormalization group arguments, 
and hence it should be valid when  $N$ is asymptotically large. 
 For the ratio $R_K^2/R^2$ we have the following  expansion: 
\begin{equation}
R_K^2/R^2 = (A_K/A)N^{2 \Delta \nu_K} (1+(B_K-B)N^{-\Delta}+O(1/N)) \, , 
\label{asympto} 
\end{equation}
where $\Delta \nu_K  = \nu_K -\nu_{\rm RW}$. 
Here we recall that $\nu_{\rm RW}=0.5$.  
For each of the four knots, we have applied the formula (\ref{asympto}) 
to the 10 data points with $N \ge 1000$ shown in Fig. 1.   
Here, when we assign the condition of $N \ge 1000$,  
 we have taken into account 
the strong finite-size effects of $R_{K}^2$ such as the 
crossover of the trivial knot.  
Another recent study \cite{Miyuki} 
on the cylinder model of SAPs \cite{PLA,PRE}   
shows that the asymptotic scaling behavior is seen only when $N > 1000$ 
for the case of very thin cylinders with the cylinder radius $r=0.001$.

\par 
The best estimates of the fitting parameters of the formula (\ref{asympto}) 
and  the $\chi^2$ values are listed in Table 2. 
From the results, 
we may conclude that the asymptotic expansion 
(\ref{asympto}) is consistent with the numerical values of 
 $R_{K}^{2}$ for $N \ge 1000$. 
 It is remarked that  the $\chi^2$ values in Table 2 are less than 
  10 for the four knots. Moreover, 
 the best estimates are compatible with several 
 different viewpoints. For instance, 
 the estimate  of $2 \Delta \nu_K$ 
 is  given by about 0.03 
  and independent of the knot type. 
This leads to an estimate of the exponent: $\nu_K \approx 0.515$.  
The value can be considered as equivalent to  
the exponent $\nu_{\rm RW}$, with respect to the errors of the analysis.  
The fact that  the exponent $\nu_K$  
is independent of the knot type 
is consistent with the interpretation on the lattice model of
Refs. \cite{Orlandini,Janse1991}. 
The estimated values of the amplitude ratio $A_K/A$ 
for the four knots 
  also seem to be independent of the knot type. 
We note that  we have assumed $\Delta$=0.5 in constructing Table 2. 
When we set $\Delta$=1.0, similar values  are obtained 
 for the best estimates of the  fitting parameters.

\par 
Let us consider a formula which  effectively  
describes the $N$-dependence 
of $R_K^2$ for $N > 2000$.  
Assuming  $\Delta \nu_K=0$ in the asymptotic formula (\ref{asympto}), 
 we have the following:  
$R_K^2/R^2 = \alpha_K (1+ \beta_K \, N^{-\Delta}+O(1/N))$.  
Here we have replaced by $\alpha_K$ 
and $\beta_K$, $A_K/A$ and $B_K -B$, respectively. 
Applying the formula  to the numerical data of $R_K^2$ of the Gaussian random 
polygon for $N \ge 1000$, 
we see that it gives good fitting curves to the data. 
The best estimates of the parameters are shown in Table 3.  
Interestingly, they are rather close to the best estimates 
for the cylinder model of SAPs  
with a very small cylinder radius,  
which are obtained by applying the same formula to 
the data of  $R_K^2$ in Ref. \cite{PRE}.  
In Table 3, the parameter $\alpha_K$ 
is roughly given by 1.5 for the Gaussian random polygon. 
On the other hand, we have the similar value for the cylinder model  
 with the cylinder radius $r = 0.001$ as shown  
 in Fig. 3 of Ref. \cite{PRE}.   
We also find in Table 3 that $\alpha_K \approx 1.5$ for the four knots. 
It  follows that the mean size $R_K$ of 
 random polygons with a specified knot $K$,  
such as the trivial, $3_1$,  $4_1$, and $3_1 \# 3_1$ knots,   
 is larger than the average size $R$ 
 of random polygons over all knots  
 in the asymptotic regime. However, 
it is consistent with the observation 
in Fig. 1 that the ratio $R_K^2/R^2$  increases monotonically 
and approaches 1.3 or 1.4 when $N \sim 2000$ (see also  Ref. \cite{PRE}).

\section{Conclusion}

\par 
We have shown  that $R_K^2$ of  Gaussian random polygons 
 have strong finite-size effects 
which should be valid for extremely large values of $N$ such as $N=2000$. 
If we remove the finite-size effects from the data analysis, 
then  the asymptotic behavior of $R_K^2$ is given by 
the standard one with the critical exponent $\nu_{\rm RW}$ 
of random walks.   Here 
we note that the main result should be valid 
 also for the numerical data of other models 
 of random polygons \cite{Miyuki,Matsuda}.  
Thus, the  studies \cite{desCloizeaux-Let,Deutsch,Grosberg} 
associated with the conjecture of the effective entropic-repulsion 
should be important for describing the strong finite-size effects 
of random polygons,  
which could appear practically in any system of ring polymers 
in solution at the theta condition.

{\vskip 1.2cm}
\par \noindent 
{\bf Acknowledgement} 

We would like to thank  Prof. K. Ito, Dr. K. Tsurusaki and 
Dr. H. Furusawa for helpful discussions. 
One of the authors (M.K.S.) would also like 
to thank Prof. A.Yu. Grosberg for helpful 
discussions.

\newpage

\begin{table}
\caption{Best estimates for the fitting lines  in  Fig.1 
describing  the anomalous scaling behavior: 
$R_K^2/R^2 = \Gamma_K \, N^{2 \Delta \nu_K^{\rm eff}}$. 
Here $\Delta \nu_{K}^{\rm eff}= \nu_K^{\rm eff}- \nu_{\rm RW}$.   
For the trivial knot, the fit is  obtained from the data points with 
$N \ge 400$. For the trefoil ($3_1$) and  figure-eight ($4_1$) knots,
 and the composite knot of $3_1 \# 3_1$, the fitting lines are obtained 
from the data for $N \ge 100$. }
\begin{ruledtabular}
\begin{tabular}{ccccc}
knot type & $\Gamma_K$ & $2 \Delta \nu_K^{\rm eff}$ & $\chi^2$ \\
\hline
triv/ave & 0.600$\pm$0.012 & 0.090$\pm$0.003 & 6\\ 
tre/ave & 0.514$\pm$0.004 & 0.121$\pm$0.001 & 14\\
$4_1$/ave & 0.431$\pm$0.006 & 0.143$\pm$0.002 & 11\\
$3_1 \sharp 3_1$/ave & 0.398$\pm$0.005 & 0.153$\pm$0.002 & 25\\
\hline
\end{tabular}
\end{ruledtabular}
\end{table}

\begin{table}
\caption{Best estimates of the fitting parameters 
of the asymptotic formula: 
$R_K^2/R^2= (A_K/A) N^{2 \Delta \nu_K} 
\left(1 + (B_K-B)N^{-\Delta} \right) $. 
Here $\Delta \nu_K = \nu_K - \nu_{\rm RW}$.} 
\begin{ruledtabular}
\begin{tabular}{cccccc}
knot type & $A_K/A$ & $B_K -B$ & $2 \Delta \nu_K $ & $\chi^2$ \\
\hline
triv/ave & 1.163$\pm$3.22 & -3.746$\pm$17.609 & 0.027$\pm$0.309 & 3\\
tre/ave & 1.131$\pm$1.837 & -5.333$\pm$9.486 & 0.033$\pm$0.183 & 9\\
$4_1$/ave & 1.129$\pm$3.837 & -5.860$\pm$19.170 & 0.033$\pm$0.385 & 7\\
$3_1 \sharp 3_1$/ave & 1.166$\pm$1.662 & -5.928$\pm$8.071 & 0.028$\pm$0.161 & 9\\
\hline
\end{tabular}
\end{ruledtabular}
\end{table}

\begin{table}
\caption{Best estimates of the fitting parameters of 
the formula  effectively describing 
the large-$N$ behavior of  the ratio: 
$R_K^2/R^2= \alpha_K \, \left(1 + \beta_K \, N^{-\Delta}  \right) $. }
\begin{ruledtabular}
\begin{tabular}{ccccc}
knot type & $\alpha_K$ & $\beta_K$ & $\chi^2$ \\
\hline
triv/ave & 1.472$\pm$0.042 & -5.187$\pm$0.833 & 3\\
tre/ave & 1.516$\pm$0.026 & -6.923$\pm$0.483 & 10\\
$4_1$/ave & 1.507$\pm$0.0056 & -7.375$\pm$1.1010 & 7\\
$3_1 \sharp 3_1$/ave  & 1.493$\pm$0.024 & -7.270$\pm$0.446 & 9\\
\hline
\end{tabular}
\end{ruledtabular}
\end{table}

\begin{figure}
\caption{Logarithmic plot of 
the ratio $R^2_K/R^2$ versus the  number $N$ of polygonal nodes of the
Gaussian random  polygon for the range from $N=100$ to $N=1900$. 
Numerical estimates of $R^2_{K}/R^2$ 
for the trivial, trefoil (31) and figure-eight (41) knots 
 are shown by black circles, black squares and black triangles, 
  respectively.
In the inset, the enlarged figure 
  shows the logarithmic plot of the estimates of $R_K^2/R^2$ from 
  $N=1000$ to 1900 for the trivial, trefoil and figure-eight knots.}
\end{figure}

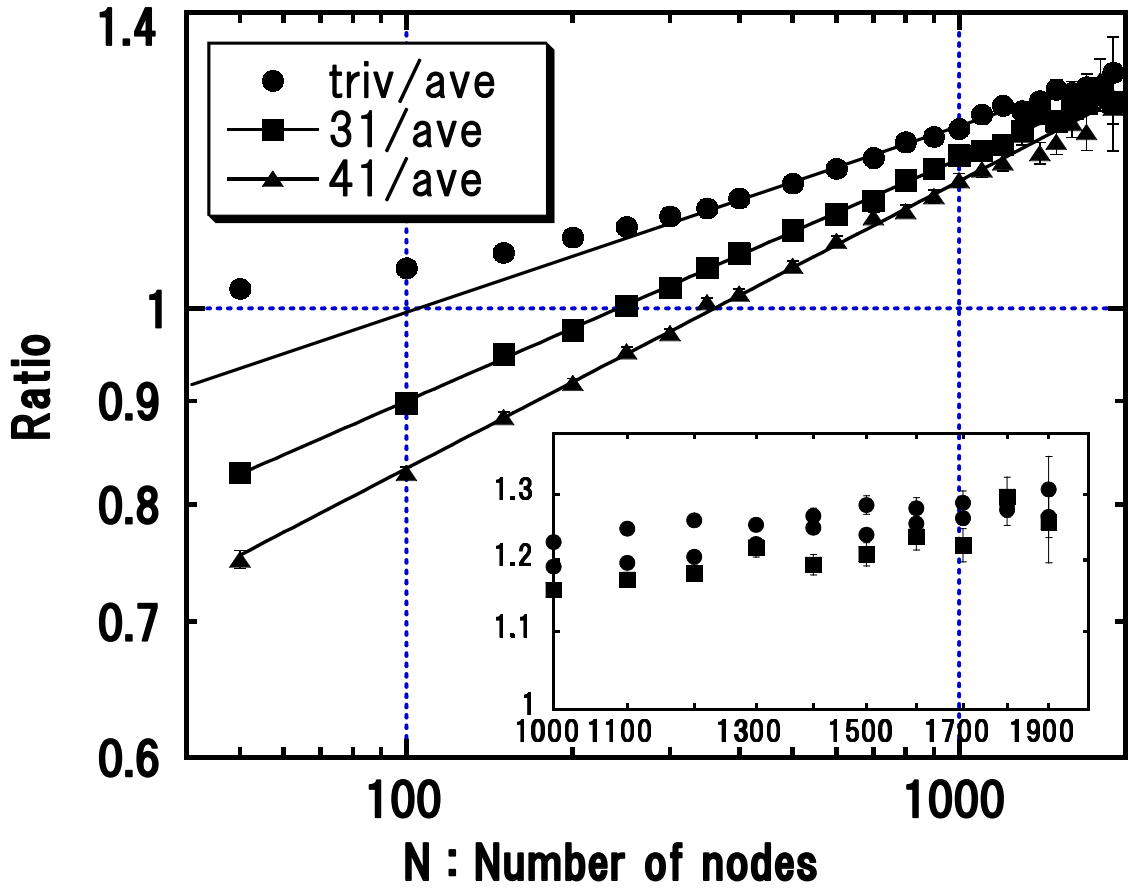
\end{document}